\documentclass[aps,prl,twocolumn,tightenlines,superscriptaddress,showpacs,byrevtex]{revtex4}
\usepackage{rotating}
\usepackage{longtable}
\usepackage{amsmath}
\usepackage{graphicx}
\usepackage{dcolumn}
\usepackage{multirow}
\usepackage{color}
\usepackage{ulem}
\usepackage{hyperref}
\RequirePackage{xspace}
\usepackage{relsize}
\def\babar{\mbox{\slshape B\kern-0.1em{\smaller A}\kern-0.1em B\kern-0.1em{\smaller A\kern-0.2em R}}}

\def\bbar   {\ensuremath{\overline b}\xspace}
\def\qqbar  {\ensuremath{q\overline q}\xspace}
\def\bbbar  {\ensuremath{b\overline b}\xspace}

\def\Kbar   {\kern 0.2em\overline{\kern -0.2em K}{}\xspace}

\def\Kz     {\ensuremath{K^0}\xspace}
\def\Kzb    {\ensuremath{\Kbar^0}\xspace}
\def\KzKzb  {\ensuremath{\Kz \kern -0.16em \Kzb}\xspace}

\def\KS     {\ensuremath{K^0_{\scriptscriptstyle S}}\xspace}

\def\Dbar   {\kern 0.2em\overline{\kern -0.2em D}{}\xspace}

\def\Dz     {\ensuremath{D^0}\xspace}
\def\Dzb    {\ensuremath{\Dbar^0}\xspace}
\def\DzDzb  {\ensuremath{\Dz {\kern -0.16em \Dzb}}\xspace}
\def\Dp     {\ensuremath{D^+}\xspace}
\def\Dm     {\ensuremath{D^-}\xspace}
\def\DpDm   {\ensuremath{\Dp {\kern -0.16em \Dm}}\xspace}

\def\Bbar   {\kern 0.18em\overline{\kern -0.18em B}{}\xspace}

\def\antiproton{\ensuremath{\overline p}\xspace}
\mathchardef\Upsilon="7107
\def\Y#1S{\ensuremath{\Upsilon{(#1S)}}\xspace}

\def\Ut     {\Y2S}

\def\Ufo    {\Y4S}
\def\Uf     {\Y5S}

\def\et     {\ensuremath{\eta_b(1S)}\xspace}
\def\ett    {\ensuremath{\eta_b(2S)}\xspace}

\def\etm    {\ensuremath{\eta_b(mS)}\xspace}

\def\hbp    {\ensuremath{h_b(2P)}\xspace}
\def\hbn    {\ensuremath{h_b(nP)}\xspace}

\def\DeltaE {\mbox{$\Delta E$}\xspace}
\def\cm   {\ensuremath{{\rm \,cm}}\xspace}
\def\invfb{\ensuremath{\mbox{\,fb}^{-1}}\xspace}

\def\to{\ensuremath{\rightarrow}\xspace}

\newcommand{\stat}{\ensuremath{\rm{(stat)}}\xspace}
\newcommand{\syst}{\ensuremath{\rm{(syst)}}\xspace}
\newcommand{\gev}{\ensuremath{\rm{\,Ge\kern -0.1em V}}\xspace}
\newcommand{\mev}{\ensuremath{\rm{\,Me\kern -0.1em V}}\xspace}
\newcommand{\gevc}{\ensuremath{{\rm{\,Ge\kern -0.1em V\!/}}{\it c}}\xspace}
\newcommand{\mevc}{\ensuremath{{\rm{\,Me\kern -0.1em V\!/}}{\it c}}\xspace}
\newcommand{\gevcc}{\ensuremath{{\rm{\,Ge\kern -0.1em V\!/}}{\it c}^2}\xspace}
\newcommand{\mevcc}{\ensuremath{{\rm{\,Me\kern -0.1em V\!/}}{\it c}^2}\xspace}

\renewcommand{\arraystretch}{1.1}

\begin{document}

\title{\boldmath Search for Bottomonium States in Exclusive Radiative $\Ut$ Decays}
\date{\today}
\begin{abstract}
  \noindent
  We search for bottomonium states in $\Ut\to (\bbbar)\gamma$ decays with an 
  integrated luminosity of $24.7\invfb$ recorded at the $\Ut$ resonance with 
  the Belle detector at KEK, containing $(157.8\pm3.6)\times10^6$ $\Ut$ events.
  The $(\bbbar)$ system is reconstructed in $26$ exclusive hadronic final 
  states composed of charged pions, kaons, protons, and $\KS$ mesons. We find 
  no evidence for the state recently observed around 9975~MeV ($X_{\bbbar}$) in 
  an analysis based on a data sample of $9.3\times 10^6$ $\Ut$ events collected
  with the CLEO~III detector. We set a $90\%$ confidence level upper limit on 
  the branching fraction
  ${\cal B}[\Ut\to X_{\bbbar} \gamma]\times\sum_i{\cal B}
  [X_{\bbbar} \to h_i]< 4.9\times 10^{-6}$, summed over the exclusive hadronic
  final states employed in our analysis.
  This result is an order of magnitude smaller than the measurement
  reported with CLEO data. We also set an upper limit for the $\et$ state of 
  ${\cal B}[\Ut\to\et\gamma]\times\sum_i{\cal B}[\et\to h_i]< 3.7\times 10^{-6}$.
\end{abstract}

\pacs{14.40.Pq, 13.25.Gv, 12.39.Pn}

\noaffiliation
\affiliation{University of the Basque Country UPV/EHU, 48080 Bilbao}
\affiliation{University of Bonn, 53115 Bonn}
\affiliation{Budker Institute of Nuclear Physics SB RAS and Novosibirsk State University, Novosibirsk 630090}
\affiliation{Faculty of Mathematics and Physics, Charles University, 121 16 Prague}
\affiliation{Chiba University, Chiba 263-8522}
\affiliation{Deutsches Elektronen--Synchrotron, 22607 Hamburg}
\affiliation{Justus-Liebig-Universit\"at Gie\ss{}en, 35392 Gie\ss{}en}
\affiliation{II. Physikalisches Institut, Georg-August-Universit\"at G\"ottingen, 37073 G\"ottingen}
\affiliation{Gyeongsang National University, Chinju 660-701}
\affiliation{Hanyang University, Seoul 133-791}
\affiliation{University of Hawaii, Honolulu, Hawaii 96822}
\affiliation{High Energy Accelerator Research Organization (KEK), Tsukuba 305-0801}
\affiliation{Hiroshima Institute of Technology, Hiroshima 731-5193}
\affiliation{Ikerbasque, 48011 Bilbao}
\affiliation{Indian Institute of Technology Guwahati, Assam 781039}
\affiliation{Indian Institute of Technology Madras, Chennai 600036}
\affiliation{Institute of High Energy Physics, Chinese Academy of Sciences, Beijing 100049}
\affiliation{Institute for High Energy Physics, Protvino 142281}
\affiliation{INFN - Sezione di Torino, 10125 Torino}
\affiliation{Institute for Theoretical and Experimental Physics, Moscow 117218}
\affiliation{J. Stefan Institute, 1000 Ljubljana}
\affiliation{Kanagawa University, Yokohama 221-8686}
\affiliation{Institut f\"ur Experimentelle Kernphysik, Karlsruher Institut f\"ur Technologie, 76131 Karlsruhe}
\affiliation{Korea Institute of Science and Technology Information, Daejeon 305-806}
\affiliation{Korea University, Seoul 136-713}
\affiliation{Kyungpook National University, Daegu 702-701}
\affiliation{\'Ecole Polytechnique F\'ed\'erale de Lausanne (EPFL), Lausanne 1015}
\affiliation{Faculty of Mathematics and Physics, University of Ljubljana, 1000 Ljubljana}
\affiliation{Luther College, Decorah, Iowa 52101}
\affiliation{University of Maribor, 2000 Maribor}
\affiliation{Max-Planck-Institut f\"ur Physik, 80805 M\"unchen}
\affiliation{School of Physics, University of Melbourne, Victoria 3010}
\affiliation{Moscow Physical Engineering Institute, Moscow 115409}
\affiliation{Moscow Institute of Physics and Technology, Moscow Region 141700}
\affiliation{Graduate School of Science, Nagoya University, Nagoya 464-8602}
\affiliation{Kobayashi-Maskawa Institute, Nagoya University, Nagoya 464-8602}
\affiliation{Nara Women's University, Nara 630-8506}
\affiliation{National United University, Miao Li 36003}
\affiliation{Department of Physics, National Taiwan University, Taipei 10617}
\affiliation{H. Niewodniczanski Institute of Nuclear Physics, Krakow 31-342}
\affiliation{Nippon Dental University, Niigata 951-8580}
\affiliation{Niigata University, Niigata 950-2181}
\affiliation{University of Nova Gorica, 5000 Nova Gorica}
\affiliation{Osaka City University, Osaka 558-8585}
\affiliation{Pacific Northwest National Laboratory, Richland, Washington 99352}
\affiliation{Panjab University, Chandigarh 160014}
\affiliation{University of Pittsburgh, Pittsburgh, Pennsylvania 15260}
\affiliation{Research Center for Electron Photon Science, Tohoku University, Sendai 980-8578}
\affiliation{University of Science and Technology of China, Hefei 230026}
\affiliation{Seoul National University, Seoul 151-742}
\affiliation{Soongsil University, Seoul 156-743}
\affiliation{Sungkyunkwan University, Suwon 440-746}
\affiliation{School of Physics, University of Sydney, New South Wales 2006}
\affiliation{Tata Institute of Fundamental Research, Mumbai 400005}
\affiliation{Excellence Cluster Universe, Technische Universit\"at M\"unchen, 85748 Garching}
\affiliation{Toho University, Funabashi 274-8510}
\affiliation{Tohoku Gakuin University, Tagajo 985-8537}
\affiliation{Tohoku University, Sendai 980-8578}
\affiliation{Department of Physics, University of Tokyo, Tokyo 113-0033}
\affiliation{Tokyo Institute of Technology, Tokyo 152-8550}
\affiliation{Tokyo Metropolitan University, Tokyo 192-0397}
\affiliation{Tokyo University of Agriculture and Technology, Tokyo 184-8588}
\affiliation{University of Torino, 10124 Torino}
\affiliation{CNP, Virginia Polytechnic Institute and State University, Blacksburg, Virginia 24061}
\affiliation{Wayne State University, Detroit, Michigan 48202}
\affiliation{Yamagata University, Yamagata 990-8560}
\affiliation{Yonsei University, Seoul 120-749}
  \author{S.~Sandilya}\affiliation{Tata Institute of Fundamental Research, Mumbai 400005} 
  \author{K.~Trabelsi}\affiliation{High Energy Accelerator Research Organization (KEK), Tsukuba 305-0801} 
  \author{G.~B.~Mohanty}\affiliation{Tata Institute of Fundamental Research, Mumbai 400005} 
  \author{I.~Adachi}\affiliation{High Energy Accelerator Research Organization (KEK), Tsukuba 305-0801} 
  \author{H.~Aihara}\affiliation{Department of Physics, University of Tokyo, Tokyo 113-0033} 
  \author{D.~M.~Asner}\affiliation{Pacific Northwest National Laboratory, Richland, Washington 99352} 
  \author{T.~Aushev}\affiliation{Institute for Theoretical and Experimental Physics, Moscow 117218} 
 \author{T.~Aziz}\affiliation{Tata Institute of Fundamental Research, Mumbai 400005} 
  \author{A.~M.~Bakich}\affiliation{School of Physics, University of Sydney, NSW 2006} 
  \author{A.~Bala}\affiliation{Panjab University, Chandigarh 160014} 
% ...
\collaboration{The Belle Collaboration}
%
%
\maketitle

{\renewcommand{\thefootnote}{\fnsymbol{footnote}}}
\setcounter{footnote}{0}

Bottomonium, a bound system of a bottom ($b$) quark and its antiquark
($\bbar$), offers a unique laboratory to study strong interactions; 
since the $b$ quark is heavier than other quarks ($q = u,d,s,c$),
the system can be described by nonrelativistic quantum mechanics and 
effective theories~\cite{Brambilla:2012cs}. Spin-singlet states 
permit the study of spin-spin interactions within the $\bbbar$ system.

The ground state of the bottomonium family with zero orbital and spin 
angular momenta, the $\et$, was discovered by the \babar\ Collaboration
in 2008~\cite{Aubert:2008ba}. Evidence for its radially excited spin-singlet
partner, the $\ett$,  was reported by the Belle 
Collaboration~\cite{Mizuk:2012pb} using a $133.4\invfb$ data sample collected 
near the $\Uf$ resonance. That analysis used the process 
$e^+e^-\to\Uf\to\hbn\pi^+\pi^-$, $h_b\to\etm\gamma$ for $n(\geq m)=1,2$. The 
$\ett$ mass measured in the $\hbp\to\ett\gamma$ transition was 
$[9999.0\pm3.5\stat^{+2.8}_{-1.9}\syst]\mevcc$, corresponding to a hyperfine mass 
splitting between $\Ut$ and $\ett$ states, 
$\Delta M_{\rm HF}(2S)\equiv M[\Ut]- M[\ett]$, of $[24.3^{+4.0}_{-4.5}]\mevcc$. 
The \babar\ and Belle analyses were based on an inclusive approach, where the 
final state of the $\eta_b(nS)$ was not reconstructed.

There is a recent claim~\cite{Dobbs:2012zn} of the observation of a bottomonium
state $X_{\bbbar}$ in the radiative decay $\Ut\to X_{\bbbar}\gamma$ with a data 
sample of $9.3\times 10^6$ $\Ut$ decays recorded with the CLEO~III detector.
The analysis, based on the reconstruction of $26$ exclusive hadronic final 
states, reports a mass of $[9974.6\pm2.3\stat\pm2.1\syst]\mevcc$ and assigns 
this state to the $\ett$, which corresponds to 
$\Delta M_{\rm HF}(2S)=[48.6\pm3.1]\mevcc$.
This disagrees with most of the predictions for $\Delta M_{\rm HF}(2S)$ from 
unquenched lattice calculations, potential models and a model-independent
relation that are compiled in Ref.~\cite{Burns} and therefore suggests a 
flaw in the theoretical understanding of QCD hyperfine mass splittings.
In contrast, the Belle result~\cite{Mizuk:2012pb} is consistent with the
theoretical expectations in Ref.~\cite{Burns}.

In this Letter, we report a search for the states $X_{\bbbar}$ in
$\Ut\to X_{\bbbar}\gamma$ decays and $\et$ in $\Ut\to \et\gamma$ decays using a 
data sample with an integrated luminosity of $24.7\invfb$ collected at the \Ut 
peak with the Belle detector~\cite{belle} at the KEKB asymmetric-energy 
$e^+e^-$ collider~\cite{KEKB}. The sample contains $(157.8\pm3.6)\times10^6$ 
$\Ut$ decays~\cite{NY2S}, which is about $17$~times larger than the one used in 
Ref.~\cite{Dobbs:2012zn}.
In addition, $1.7$ $[89.5]\invfb$ of data recorded $30$ $[60]\mev$ below the 
$\Ut$ $[\Ufo]$ resonance energy (``off-resonance'') are used to model the 
$e^+e^-\to\qqbar$ continuum background. It is not possible to reconstruct the 
$\ett$ state using exclusive reconstruction of the hadronic final state near 
the mass found in Ref.~\cite{Mizuk:2012pb} because this region suffers from a 
low photon detection efficiency and high background.

We employ the \textsc{EvtGen}~\cite{evtgen} package to generate signal 
Monte Carlo (MC) events. The radiative decays of the $\Ut$ are generated using
the helicity amplitude formalism~\cite{helamp}. Hadronic decays of the 
$(\bbbar)$ system are modeled assuming a phase space distribution; to 
incorporate final state radiation effects, an interface to 
\textsc{Photos}~\cite{photos} is added. Inclusive $\Ut$ MC events, produced 
using \textsc{Pythia}~\cite{pythia} with the same luminosity as the data, are 
investigated for potential peaking backgrounds. 

The Belle detector~\cite{belle} is a large-solid-angle spectrometer that 
includes a silicon vertex detector, a 50-layer central drift chamber (CDC),
an array of aerogel threshold Cherenkov counters (ACC), time-of-flight
scintillation counters (TOF), and an electromagnetic calorimeter (ECL) 
comprising CsI(Tl) crystals. All these components are located inside a 
superconducting solenoid coil that provides a $1.5$\,T magnetic field.

Our event reconstruction begins with the selection of an appropriate number 
and type of charged particles to reconstruct a subset of the many exclusive
hadronic final states of the $(\bbbar)$ system. We restrict ourselves to the 
$26$ modes reported in Ref.~\cite{Dobbs:2012zn}: 
$2(\pi^+\pi^-)$, $3(\pi^+\pi^-)$,
$4(\pi^+\pi^-)$, $5(\pi^+\pi^-)$, $K^+K^-\pi^+\pi^-$, $K^+K^-2(\pi^+\pi^-)$,
$K^+K^-3(\pi^+\pi^-)$, $K^+K^-4(\pi^+\pi^-)$, $2(K^+K^-)$, $2(K^+K^-)\pi^+\pi^-$,
$2(K^+K^-\pi^+\pi^-)$, $2(K^+K^-)3(\pi^+\pi^-)$, $\pi^+\pi^-p\antiproton$,
$2(\pi^+\pi^-)p\antiproton$, $3(\pi^+\pi^-)p\antiproton$, $4(\pi^+\pi^-)p
\antiproton$, $\pi^+\pi^-K^+K^-p\antiproton$, $2(\pi^+\pi^-)K^+K^-p\antiproton$,
$3(\pi^+\pi^-)K^+K^-p\antiproton$, $\KS K^{\pm}\pi^{\mp}$, $\KS K^{\pm}\pi^{\mp}
\pi^+\pi^-$, $\KS K^{\pm}\pi^{\mp}2(\pi^+\pi^-)$, $\KS K^{\pm}\pi^{\mp}
3(\pi^+\pi^-)$, $2\KS(\pi^+\pi^-)$, $2\KS 2(\pi^+\pi^-)$, and 
$2\KS 3(\pi^+\pi^-)$.

We require all charged tracks, except for those from $\KS$ decays, to
originate from the vicinity of the interaction point (IP) by requiring 
their impact parameters along and perpendicular to the $z$ axis 
to be less than $4$ and $1\cm$, respectively. Here, the $z$ axis is
defined by the direction opposite the $e^+$ beam. Track candidates are
identified as pions, kaons, or protons (``hadrons'') based on information
from the CDC, the TOF and the ACC. The kaon identification efficiency
is $83\% - 91\%$ with a pion misidentification 
probability of $8\% - 10\%$. Pions are detected with an efficiency of
$87\% - 89\%$ with  a kaon-to-pion misidentification rate of  $7\% - 13\%$. The
proton identification efficiency is $95\%$, while the probability of a
kaon being misidentified as a proton is below $3\%$. Candidate $\KS$
mesons are reconstructed by combining two oppositely charged tracks (with
a pion mass assumed for both) with an invariant mass between $486$ and 
$509$ $\mevcc$;
the selected candidates are also required to satisfy the criteria
described in Ref.~\cite{Ks_selection} to ensure that their decay
vertices are displaced from the IP.

We then combine a photon candidate with the $(\bbbar)$ system to form an 
$\Ut$ candidate. The photon is reconstructed from an isolated
(not matched to any charged track) cluster in the ECL that has
an energy greater than $22\mev$ and a cluster shape consistent with an 
electromagnetic shower: the energy sum of the $3\times3$
array of crystals centered around the most energetic one exceeding $85\%$
of that of the $5\times5$ array of crystals. 
The energy of the signal photon is  $30-70$ MeV and
$400-900$ MeV for the $X_{\bbbar}$ and \et, respectively. We exclude photons 
from the backward endcap in the $\et$ selection to suppress low-energy photons
arising from beam-related background. For the $X_{\bbbar}$ selection, 
both the backward and forward endcap regions are excluded as the energy of the
photon from the $\Ut\to X_{\bbbar} \gamma$ decay is too low, and lies in a range 
contaminated with large beam backgrounds. The photon energy resolution in the 
barrel ECL ranges between
$2\%$ at $E_{\gamma}=1\gev$ and $3\%$ at $E_{\gamma}=100\mev$.

There is a weak correlation between the signal photon
momentum and the thrust axis of the hadrons of the $(\bbbar)$ system if the
latter has spin zero. The same correlation 
is stronger for continuum events~\cite{Aubert:2008ba}, so 
the cosine of the angle $\theta_T$ between the candidate photon and the 
thrust axis, calculated in the $e^+e^-$ center-of-mass (CM) frame, is useful in 
suppressing the continuum background. Since the  distribution of this variable 
is independent of the ($\bbbar$)-mass region considered, we require 
$|\cos\theta_T|<0.8$ for a substantial reduction (60\%) of continuum events 
and a modest loss (20\%) of signal.

The signal windows for the difference between the energy of the $\Ut$ candidate
and the CM energy ($\DeltaE$) and the $\Ut$ momentum measured in the CM frame 
($P_{\Ut}^*$) are optimized separately for the $X_{\bbbar}$ and $\et$ mass
regions. We perform this optimization using a figure-of-merit $S/\sqrt{S+B}$,
where $S$ is the expected signal based on MC simulations, and $B$ is the
background estimated from a sum of the $\Ufo$ off-resonance data, scaled to the
available $\Ut$ integrated luminosity, and the inclusive $\Ut$ MC sample
described earlier. The value of $S$ is calculated by assuming
the branching fraction to be $46.2\times10^{-6}$ for the 
$X_{\bbbar}$~\cite{Dobbs:2012zn} and $3.9\times10^{-6}$ for the 
$\et$~\cite{Aubert:2009as}. 
The $\Ut$ candidates with $-40\mev<\DeltaE<50\mev$ and $P_{\Ut}^*<30\mevc$
[$-30\mev<\DeltaE<80\mev$ and $P_{\Ut}^*<50\mevc$] are
retained for a further study of the $X_{\bbbar}$ [$\et$] state.
For the two-body decay hypothesis, the angle $\theta_{(\bbbar)\gamma}$ between
the reconstructed ($\bbbar$) system and the photon candidate in the
CM frame should be close to $180^\circ$. We apply an optimized requirement 
on $\theta_{(\bbbar)\gamma}$ to be greater than $150^\circ$ [$177^\circ$] 
to select the $\Ut\to X_{\bbbar} \gamma$ [$\Ut\to \et \gamma$]
decay candidates. The difference between the invariant mass formed by 
combining the signal photon with another photon candidate in the event and
the nominal $\pi^0$ mass~\cite{pdg2012} is computed for each photon pair;
the smallest of the magnitudes of these differences is denoted by 
$\Delta M_{\gamma\gamma}$ and used for a $\pi^0$ veto.
For the $\et$ selection, where the background contribution is dominated by 
$\pi^0$'s coming from the $\Ut$ decays, we require
$\Delta M_{\gamma\gamma}>10\mevcc$.
 We do not apply the $\pi^0$ veto in the $X_{\bbbar}$ selection since there is
negligible $\pi^0$ contamination; the background here is dominated by photons 
coming from beam background.
The final selection efficiencies for the individual modes range from $6.1\%$ 
[$X_{\bbbar}\to 3(\pi^+\pi^-)$] to $1.2\%$ [$X_{\bbbar}\to 2\KS 3(\pi^+\pi^-)$].

We apply a kinematic fit to the $\Ut$ candidates constrained by
energy-momentum conservation. The resolution of the reconstructed invariant mass
of the $\et$, presented in terms of 
$\Delta M\equiv M[(\bbbar)\gamma]-M(\bbbar)$,
is significantly improved by this fit from approximately $14$ to $8\mevcc$. 
The improvement in the mass resolution is minimal for the $X_{\bbbar}$ since
the photon has so little energy. The fit $\chi^2$ value is used to select the 
best $\Ut$ candidate in the case of multiple candidates that appear in about 
$10\%$ of the events satisfying the $X_{\bbbar}$ selection. 

We extract the signal yield by performing an unbinned extended 
maximum-likelihood fit to the $\Delta M$ distribution for all selected 
candidates. The probability density functions (PDFs) for $\chi_{bJ}(1P)$ and 
$X_{\bbbar}$ signals are parametrized by the sum of a Gaussian and an
asymmetric Gaussian function to take into account low-energy tails. 
Their parameters (the common mean, three widths, and the relative fraction) 
are taken from MC simulations. To account for the modest difference in 
the detector resolution between data and simulations, we use a 
calibration factor common to the four signal components, {\it i.e.}, 
$\chi_{bJ}(1P)$ with $J=0,1,2$ and $X_{\bbbar}$, to smear their core Gaussian 
components. 
\begin{figure}[htbp]
\includegraphics[width=7.9cm]{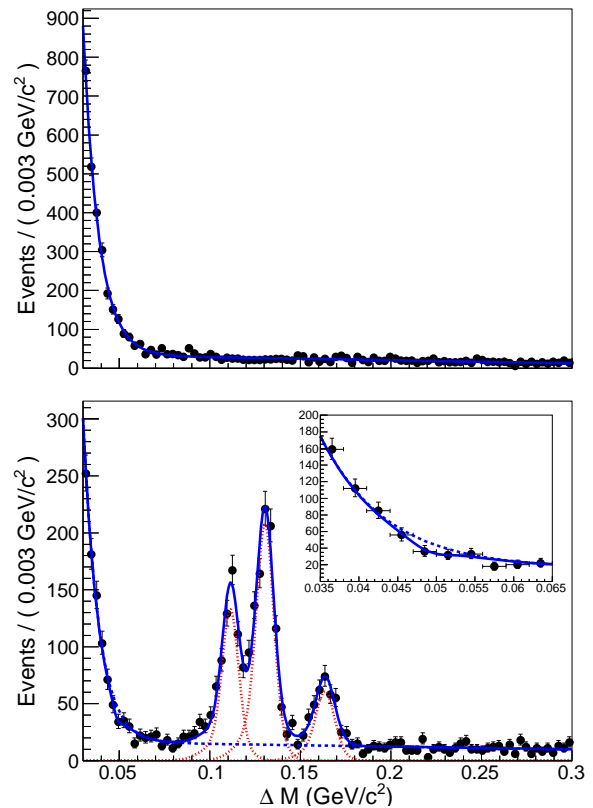}
\caption{(color online). The $\Delta M$ distributions  for (top) $\Ufo$ 
  off-resonance data and (bottom) $\Ut$ data events that pass the selection
criteria applied for the $[0.03,0.30]\gevcc$ region. Points with error bars
are the data, (top) the blue solid curve is the result of the fit
for the background-only hypothesis, and (bottom) the result of the fit for
the signal-plus-background hypothesis, where blue solid and blue dashed
curves are total fit and background components, respectively. The three
$\chi_{bJ}(1P)$ components indicated by the red dotted curves are here
considered as part of the signal. The bottom inset shows an expanded
view of the $\Delta M$ distribution in the $[0.035,0.065]\gevcc$ region.}
\label{fig:etab_2S}
\end{figure}
The choice of the background PDF is particularly important and is determined
from the large sample  of $\Ufo$ off-resonance data. 
As shown in the top plot of Fig.~\ref{fig:etab_2S}, the best fit to these data 
is obtained by using a sum of an exponential function and 
a first-order Chebyshev polynomial for the $X_{\bbbar}$
region, whose parameters are allowed to vary in the fit. This is in 
contrast to Ref.~\cite{Dobbs:2012zn}, 
where a single exponential function was used to 
describe the background PDF. 
The polynomial component is needed to model the background due 
to final-state radiation 
for $\Delta M$ $<0.15\gevcc$ and from $\pi^0$ for 
$\Delta M$ $\ge 0.15\gevcc$. We have verified using a large number of
pseudoexperiments that if the $X_{\bbbar}$ signal is present in our data sample
we would observe it with a significance above 10 standard deviations.

In the bottom plot of Fig.~\ref{fig:etab_2S}, we present 
fits to the  $\Delta M$ distributions for the sum of the $26$ modes in the 
$X_{\bbbar}$ region. The results of the fit show no evidence of an $X_{\bbbar}$ 
signal, with a yield of $-30 \pm 19$ events.
In the fits to the $\chi_{bJ}(1P)$ ($J=0,1,2$) states we observe
large signal yields and determine invariant masses of
$9859.6\pm 0.5$, $9892.8\pm 0.2$ and $9912.0\pm 0.3\mevcc$, respectively, 
which are in excellent agreement with the corresponding world-average
values~\cite{pdg2012}. The strong $\chi_{bJ}(1P)$ signals determine 
the aforementioned data-MC width-calibration factor to be
$1.23\pm 0.05$. The parameters obtained for the background PDF in the $\Ut$ 
sample are consistent with those found in the fit to the 
$\Ufo$ off-resonance
data, giving us confidence in our background modeling.

\begin{figure}[htbp]
\includegraphics[width=7.9cm]{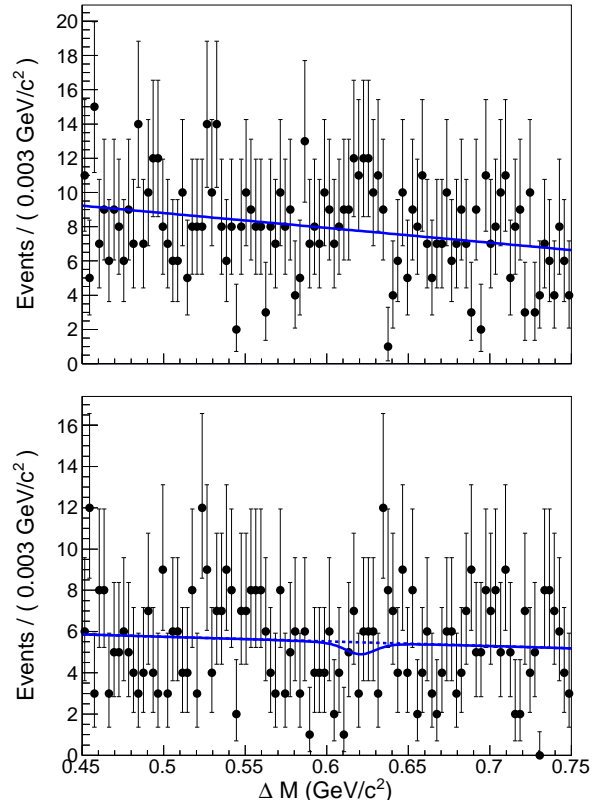}
\caption{(color online). The $\Delta M$ distributions for (top) $\Ufo$ 
  off-resonance data and (bottom) $\Ut$ data events that pass the selection 
  criteria applied for the $[0.45,0.75]\gevcc$ region. Points with error bars 
  are the data, (top) the blue solid curve is the result of the fit for the 
  background-only hypothesis, and (bottom) the result of the fit for the 
  signal-plus-background hypothesis, where blue solid and blue dashed curves
  are total fit and background components, respectively.}
\label{fig:etab_1S}
\end{figure}

The signal PDF for the $\et$ is a Breit-Wigner function, whose width is fixed 
to the value obtained in Ref.~\cite{Mizuk:2012pb}, convolved with a Gaussian
function with a width of $8\mevcc$ describing the detector resolution. 
A first-order Chebyshev polynomial is used for the background in the $\et$ 
region, validated with the large sample of $\Ufo$ off-resonance data. 
The result of the fit to off-resonance data is presented in the top plot
of Fig.~\ref{fig:etab_1S}. No signal ($-6\pm 10$ events) is found for the 
$\et$, as shown in the bottom plot of Fig.~\ref{fig:etab_1S}.

For a particle of mass near $10\gevcc$, exclusive decays are distributed across
many final states, and thus we use the $\chi_{b0}(1P)$ 
[spin-zero, as for the $\eta_b(nS)$] decay modes for guidance. The average 
efficiency for each $(\bbbar)$ state is calculated with the individual 
efficiencies [$\varepsilon_{(\bbbar)}^{i}$] obtained with MC samples weighted 
according to the yields [$N_{\chi_{b0}(1P)}^{i}$] for each mode in the 
$\chi_{b0}(1P)$ case, as 
\begin{eqnarray}              
\varepsilon[(\bbbar)]=\sum\limits_{i=1}^{26}\frac{\varepsilon_{(\bbbar)}^{i}
\times N_{\chi_{b0}(1P)}^{i}}{N_{\chi_{b0}(1P)}^{\rm tot}},
\label{eq:eff}
\end{eqnarray}
where $N_{\chi_{b0}(1P)}^{\rm tot}$ denotes the total sum of the signal yields
obtained for the $26$ hadronic decays of the $\chi_{b0}(1P)$. 
Those efficiencies are corrected to take into account the data-MC difference 
in the hadron identification efficiency. The corrected efficiencies  
are 2.9\% and 3.5\% for the $X_{\bbbar}$ and \et, respectively.
Very similar results are obtained when
using the $\chi_{b1}(1P)$ or $\chi_{b2}(1P)$ state as the proxy instead of the 
$\chi_{b0}(1P)$.

We estimate the uncertainties on the signal yields due to the signal
PDF shapes using $\pm 1\sigma$ variations of the shape
parameters that are fixed in the fit. 
The dominant sources of such additive systematic errors are the 
$X_{\bbbar}$~\cite{Dobbs:2012zn} and $\et$~\cite{Mizuk:2012pb} masses.
For the upper limit estimates (described below), we conservatively
use the fit likelihood, which gives the largest upward variation of
the signal yield: $18$ and $4$ events for the $X_{\bbbar}$ and $\et$,
respectively. The multiplicative systematic uncertainties
that do not affect the signal yields are
summarized in Table~\ref{tab:syst}. The largest contribution arises
from the uncertainty in the efficiency estimate. Two sources
dominate here: (a) the statistical error in the yield of the 
different decay modes of the $\chi_{b0}(1P)$ and 
(b) the effects of possible intermediate states on the signal efficiency 
(referred to as ``decay modeling''). 
As described earlier, all our signal MC samples are generated
with a phase space distribution. Therefore, in order to
estimate the contribution from source (b), possible intermediate states such as
$\rho^0\to\pi^+\pi^-$, $K^*(892)^0\to K^+\pi^-$ and $K^*(892)^{\pm}\to
\KS\pi^{\pm}$ are considered. Differences in the efficiencies based
on the same final-state modes generated with these intermediate 
resonances can be as large as $9.2\%$. The other minor sources arise from hadron
identification, charged track reconstruction, $\KS$ and photon detection, 
and the number of $\Ut$. 

\begin{table}[tbh]
\caption{Multiplicative systematic uncertainties (in $\%$) considered in the 
estimation of the $X_{\bbbar}$ and $\et$ upper limits. }
\label{tab:syst}
\renewcommand{\arraystretch}{1.1}
\begin{ruledtabular}
\begin{tabular}{l|cc}
  Source & $X_{\bbbar}$ & $\et$ \\
\hline
Efficiency calculation     & $\pm 2.5$ & $\pm 2.9$  \\
Decay modeling             & $\pm 9.2$ & $\pm 6.9$  \\
Hadron identification      & $\pm 3.7$ & $\pm 3.7$  \\
Track reconstruction       & $\pm 2.6$ & $\pm 2.6$  \\
$\KS$ detection            & $\pm 0.2$ & $\pm 0.2$  \\
Photon detection           & $\pm 3.0$ & $\pm 3.0$  \\
Number of $\Ut$            & $\pm 2.3$ & $\pm 2.3$  \\
\hline
Total                      & $\pm 11.2$ & $\pm 9.5$ \\
\end{tabular}
\end{ruledtabular}
\end{table}

The branching fraction is determined from the number of observed
signal events ($n_{\rm sig}$) as ${\cal B}=n_{\rm sig}/\{\varepsilon[(\bbbar)]
\times N_{\Ut}\}$, where $\varepsilon[(\bbbar)]$ is evaluated according to 
Eq.~(\ref{eq:eff}) and $N_{\Ut}$ is the total number of $\Ut$ decays.
In the absence of the signal, we obtain an upper limit at $90\%$ confidence 
level (C.L.) on the branching
fraction ($\cal{B}_{\rm UL}$) by integrating the likelihood ($\cal{L}$)
of the fit with fixed values of the branching fraction: 
$\int_{0}^{\cal{B}_{\rm UL}} {\cal{L}}({\cal{B}}) d{\cal{B}}
=0.9\times \int_0^{1}{\cal{L}}({\cal{B}}) d{\cal{B}}$. 
Multiplicative systematic uncertainties are included by convolving the 
likelihood function with a Gaussian function with a width equal to the total 
uncertainty. We estimate  
${\cal B}[\Ut\to\et\gamma]\times\sum_i{\cal B}[\et\to h_i]< 3.7\times 10^{-6}$ 
and 
${\cal B}[\Ut\to X_{\bbbar}\gamma]\times\sum_i{\cal B}[X_{\bbbar}\to h_i]< 
4.9\times 10^{-6}$.

In summary, we have searched for the $X_{\bbbar}$ state reported 
in Ref.~\cite{Dobbs:2012zn}, that is reconstructed in $26$ exclusive hadronic
final states using a sample of $(157.8\pm3.6)\times10^6$
$\Ut$ decays. We find no evidence for a signal and thus determine 
  a 90\% C.L. upper limit on the product branching fraction
${\cal B}[\Ut\to X_{\bbbar}\gamma]\times\sum_i{\cal B}
[X_{\bbbar}\to h_i]< 4.9\times 10^{-6}$, which is an order of magnitude smaller 
than the branching fraction reported in Ref.~\cite{Dobbs:2012zn}.
We have also searched for the $\et$ state and set an upper limit
${\cal B}[\Ut\to\et\gamma]\times\sum_i{\cal B}[\et\to h_i]< 3.7
\times 10^{-6}$ at 90\% C.L.

We thank the KEKB group for excellent operation of the
accelerator; the KEK cryogenics group for efficient solenoid
operations; and the KEK computer group, the NII, and 
PNNL/EMSL for valuable computing and SINET4 network support.  
We acknowledge support from MEXT, JSPS and Nagoya's TLPRC (Japan);
ARC and DIISR (Australia); FWF (Austria); NSFC (China); MSMT (Czechia);
CZF, DFG, and VS (Germany);
DST (India); INFN (Italy); MEST, NRF, GSDC of KISTI, and WCU (Korea); 
MNiSW and NCN (Poland); MES and RFAAE (Russia); ARRS (Slovenia);
IKERBASQUE and UPV/EHU (Spain); 
SNSF (Switzerland); NSC and MOE (Taiwan); and DOE and NSF (U.S.).


\begin{thebibliography}{99}

\bibitem{Brambilla:2012cs}
N. Brambilla {\it et al.},
Eur. Phys. J. C {\bf 71}, 1534 (2011).

\bibitem{Aubert:2008ba}
 B. Aubert {\it et al.} (\babar\ Collaboration),
 Phys.\ Rev.\ Lett.\ {\bf 101}, 071801 (2008); {\bf 102}, 029901(E) (2009).

\bibitem{Mizuk:2012pb}
 R. Mizuk {\it et al.} (Belle Collaboration),
 Phys.\ Rev.\ Lett.\ {\bf 109}, 232002 (2012).

\bibitem{Dobbs:2012zn}
 S. Dobbs, Z. Metreveli, A. Tomaradze, T. Xiao and K.K. Seth,
 Phys.\ Rev.\ Lett.\ {\bf 109}, 082001 (2012).

\bibitem{Burns}
 T.~J. Burns, Phys.\ Rev.\  D {\bf 87}, 034022 (2013).

\bibitem{belle}
 A. Abashian {\it et al.} (Belle Collaboration), Nucl. Instrum. Methods 
 Phys. Res., Sect. A {\bf 479}, 117 (2002); also, see the detector section
 in J. Brodzicka {\it et al.}, Prog. Theor. Exp. Phys.04D001 (2012).

\bibitem{KEKB}
 S. Kurokawa and E. Kikutani, Nucl. Instrum. Methods Phys. Res., Sect.
 A {\bf 499}, 1 (2003), and other papers included in this volume;
 T. Abe {\it et al.}, Prog. Theor. Exp. Phys., 03A001 (2013) and
 following articles up to 03A011.

\bibitem{NY2S}
X. L. Wang {\it et al.} (Belle Collaboration), 
Phys.\ Rev.\  D {\bf 84}, 071107(R) (2011).

\bibitem{evtgen}
 D. J. Lange,
 Nucl. Instrum. Methods Phys. Res., Sect. A {\bf  462}, 152 (2001).

\bibitem{helamp}
 M. Jacob and G. C. Wick,
 Ann. Phys. (N.Y.) {\bf 7}, 404 (1959); Ann. Phys. (N.Y.) {\bf 281}, 774 (2000).

\bibitem{photos}
 E. Barberio and Z. W\c{a}s, Comput. Phys. Commun. {\bf 79}, 291 (1994);
 P. Golonka and Z. W\c{a}s, Eur. Phys. J. C {\bf 45}, 97 (2006); {\bf 50}, 
 53 (2007).

\bibitem{pythia}
 T. Sj$\ddot{\rm o}$strand, S. Mrenna, and P. Skands,
 Comput.\ Phys.\ Commun.\ {\bf 178}, 852 (2008).

\bibitem{Ks_selection}
 K.-F. Chen {\it et al.} (Belle Collaboration),
 Phys.\ Rev.\ D {\bf 72}, 012004 (2005).

\bibitem{Aubert:2009as}
 B. Aubert {\it et al.} (\babar\ Collaboration),
 Phys.\ Rev.\ Lett.\ {\bf 103}, 161801 (2009).

\bibitem{pdg2012}
 J. Beringer {\it et al.} (Particle Data Group),
 Phys.\ Rev.\ D {\bf 86}, 010001 (2012).

\end{thebibliography}
\end{document}